\documentclass[ preprint,useAMS,usenatbib]{mn2e}
\usepackage{epsf}
\usepackage{graphicx}
\usepackage{amsmath,amssymb,amsbsy,amsfonts}

\newcommand{\be}{\begin{equation}}
\newcommand{\ee}{\end{equation}}
\newcommand{\bear}{\begin{eqnarray}}
\newcommand{\eear}{\end{eqnarray}}

\newcommand{\x}{{\rm x}}

\newcommand{\n}{{\rm n}}
\newcommand{\N}{{\rm c}}
\newcommand{\p}{{\rm p}}

\newcommand{\Dp}{\Delta_\N}

\newcommand{\en}{{\varepsilon_\n}}
\newcommand{\ec}{{\varepsilon_\N}}

\newcommand{\vA}{v_{\rm A}}
\newcommand{\B}{\hat{B}}
\newcommand{\cA}{c_{\rm A}}

\begin{document}

\title{Superfluid signatures in magnetar seismology}

\author[Andersson, Glampedakis \& Samuelsson]{N. Andersson$^1$,
K. Glampedakis$^{2,3}$ \& L. Samuelsson$^4$
\\
$^1$ School of Mathematics, University of Southampton,
Southampton SO17 1BJ, United Kingdom\\
$^2$ SISSA, via Beirut 2-4, 34014 Trieste, Italy\\
$^3$ Theoretical Astrophysics, Auf der Morgenstelle 10, University of Tuebingen, Tuebingen D-72076, Germany \\
$^4$Nordita, Roslagstullsbacken 23, 106 91 Stockholm, Sweden}

\maketitle

\begin{abstract}
We investigate the role of neutron star superfluidity for magnetar oscillations.
Using a plane-wave analysis we estimate the effects of a
neutron superfluid in the elastic crust region. We demonstrate that the superfluid
imprint is likely to be more significant than the effects of the crustal magnetic field.
We also consider the region immediately beneath the crust, where superfluid neutrons are thought
to coexist with a type~II proton superconductor. Since the magnetic field in the latter is
carried by an array of fluxtubes, the dynamics of this region differs from standard magnetohydrodynamics.
We show that the presence of the neutron superfluid (again) leaves a clear imprint on the oscillations of the system.
Taken together, our estimates  show that the superfluid components cannot be ignored
in efforts to carry out ``magnetar seismology''. This increases the level of complexity of the
modelling problem, but also points to the possibility of using observations to probe the superfluid nature of
supranuclear matter.
\end{abstract}


\section{Introduction}

Anomalous X-ray pulsars (AXPs) and Soft Gamma-ray Repeaters (SGRs) are widely
believed to be magnetars; neutron stars powered by an ultrastrong magnetic field (see \citet{woods} for a review).
Observations
(mainly conducted by X-ray satellites) have established basic parameters like the
magnetic field intensity, $ B \sim 10^{15}\, {\rm G} $,
and spin period, $P \sim 10 {\rm s} $, for this class of objects.
They have also revealed a complex emission pattern with
alternating periods of burst activity and quiescence. SGRs, which are typically more active than AXPs,  are
the only ones exhibiting giant flares. These flares, which are
believed to be triggered by some sort of instability in the magnetic field \citep{DT,TD},  are by far
the most energetic events
associated with magnetars.

An exciting contribution to magnetar phenomenology was provided by the recent discovery
of quasi-periodic oscillations (QPOs) in the late tail
spectrum of the two giant flares  \citep{israel,anna1,anna2}. There may also be evidence for a single QPO in the data of the third
known flare, observed back in the 1970s \citep{barat}. The frequencies of the most prominent QPOs
lie in the range $30-100\,{\rm Hz} $, exactly
where one would expect to find the seismic oscillation modes of the magnetar's crust \citep{hansen}. This
is consistent with the theoretical expectation that the energy released in a giant flare  is sufficient to
fracture the crust and excite its normal modes \citep{duncan}. If this interpretation of the QPOs is correct
then we may have the opportunity to carry out magnetar "asteroseismology"; a
comparison between theoretically predicted mode frequencies and the QPO data,
with the ultimate goal of constraining the properties of neutron star matter \citep{lars}.

Indeed, several recent papers have attempted to constrain the bulk equation of state of neutron star matter, assuming
that the observed QPO frequencies are identified with the first few toroidal seismic modes of the crust  (see \citet{stroh_review} for a
recent review and  references). This is
natural as a first step, but
in reality the situation is likely to be more complicated. As suggested by \citet{levin06} and \citet{toypaper} the strong
magnetic field would
likely couple an oscillating crust to the liquid core on a very short timescale. Then the observed
QPOs would be the manifestation of the coupled crust-core dynamics rather than the dynamics of the crust alone.
Possible evidence that the magnetic core plays an active role is given by the presence of a
low frequency QPO in the data of the December 2004 flare in SGR 1806-20. This QPO is
difficult to reconcile with the seismic mode interpretation \citep{israel}. It is therefore conceivable that
 magnetar ``seismology''  also probes the (much less well known) properties of the
interior magnetic field. This obviously comes at a price. We now have to model global crust-core oscillations,
a problem that is considerably more challenging than that of pure seismic crust modes.

Another aspect of neutron star physics, which is directly relevant to the mode interpretation of the QPOs,
has received almost no attention so far. Young and mature neutron stars (older than a month or so)
are sufficiently cold that the bulk of their interior liquid matter is in a superfluid state.
In the crust, for densities
above the nuclear drip density $ \rho \approx 4 \times 10^{11}\,{\rm g}/{\rm cm}^3 $, the  "dripped"
neutrons are expected
to form a superfluid below a temperature $\sim  5 \times 10^{9}\, {\rm K} $. The dynamical role of these
"free" neutrons could be important. After all, they account for $\sim 80\%$ of the total mass in the crust.
Similarly, in the liquid core  we  expect to
find both neutrons and protons in a superfluid state (below a similar threshold temperature).
In the outer core, the protons most likely form a type II superconductor, provided that
the interior magnetic field does not exceed a critical value
$\approx 3 \times 10^{16}\, {\rm G} $. As a consequence, any magnetic field that penetrates the proton plasma will form
a large number of quantised magnetic fluxtubes \citep{baym}.

It is clearly relevant to ask to what extent  the physical components (the crust and the magnetic field)
that play the leading role in the magnetar QPO problem are sensitive to neutron and proton superfluidity/superconductivity.
The aim of this investigation is to provide an insight into this issue, and improve our understanding of the relative
importance of the multifluid aspects of the magnetar oscillation problem. By carrying out a local analysis, i.e.
considering uniform parameter models, we learn how the shear waves in the neutron star crust are affected
by the presence of a superfluid neutron component. Similarly, a local analysis in the core tells us
how the Alfv\'en waves are altered by the presence of the neutron superfluid (which provides the bulk
of the core mass). Not surprisingly, the entrainment between neutrons and protons turns out to be the key parameter
in these problems. This initial (order of magnitude)
analysis  serves as a useful guideline for future (more detailed) work for realistic neutron star models.


\section{Multifluid dynamics of the crust}

\subsection{Lagrangian perturbation equations}

We want to model linear perturbations in a neutron star crust penetrated by a superfluid neutron component.
It is natural to use a Lagrangian picture to describe this problem. Hence, we combine the two-fluid Lagrangian perturbation equations \citep{kirsty}
with the relevant elastic terms and the magnetic force \citep{magnetcfs}. Since all known magnetars are slowly rotating, with periods of
several seconds, it makes sense to focus on the non-rotating problem. Then we need an  equation for the superfluid
neutron displacement which can be written
\begin{multline}
(1- \en) \partial_t^2 \xi^\n_i + \en  \partial_t^2 \xi^\N_i + \nabla_i \delta \Phi + \xi_\n^j \nabla_j \nabla_i \Phi \\
- (\nabla_i \xi_\n^j) \nabla_j \tilde{\mu}_\n + \nabla_i \Delta_\n \tilde{\mu}_\n  = 0
\label{eulern1}
\end{multline}
We label the variables associated with the neutrons with a constituent index n.
 $\Delta_\n$ represents a Lagrangian variation along the neutron flow (associated with the displacement $\xi_\n^i$). 
The variable $\en$ (assumed constant here) encodes the entrainment effect,  $\delta \Phi$
represents the (Eulerian) perturbation of the  gravitational potential $\Phi$ and $\tilde{\mu}_\n$ is the (specific) chemical 
potential for the neutrons.
We have
\begin{multline}
\Delta_\n \tilde{\mu}_\n = \left( { \partial \tilde{\mu}_\n \over \partial n_\n} \right)_{n_\N} \nabla_i (n_\n \xi_\n^i) \\
+ \left( { \partial \tilde{\mu}_\n \over \partial n_\N} \right)_{n_\n} \nabla_i (n_\N \xi_\N^i) + \xi_\n^i \nabla_i \tilde{\mu}_\n
\end{multline}
where $n_\n$ and $n_\N$ are the number densities of the neutrons and the baryons making up the crust nuclei, respectively.
It should be noted that we are not accounting for effects due to the presence of neutron vortices, e.g. the
mutual friction and the vortex tension here. In principle, these effects will be present even in the very
slowly rotating magnetars, and it will be interesting to consider them at a later stage. Our initial aim  
is to explore the leading order effects of this rather complicated problem.

The corresponding equation of motion for the crust nuclei can, labeling the variables associated with the nuclei with the constituent index c, be written
\begin{multline}
 (1- \ec) \partial_t^2 \xi^\N_i + \ec  \partial_t^2 \xi^\n_i + \nabla_i \delta \Phi + \xi_\N^j \nabla_j \nabla_i \Phi \\
- (\nabla_i \xi_\N^j) \nabla_j \tilde{\mu}_\N + \nabla_i \Delta_\N \tilde{\mu}_\N  = \Delta_\N f_i^\mathrm{el} + \Delta_\N f_i^\mathrm{mag}
\label{eulerp1}\end{multline}
Here $\Delta_\N$ representing the Lagrangian variation along the crust motion (associated with a displacement $\xi_\N^i$), and
\begin{multline}
\Delta_\N \tilde{\mu}_\N = \left( { \partial \tilde{\mu}_\N \over \partial n_\N} \right)_{n_\n} \nabla_i (n_\N \xi_\N^i) \\
+ \left( { \partial \tilde{\mu}_\N \over \partial n_\n} \right)_{n_\N} \nabla_i (n_\n \xi_\n^i) + \xi_\N^i \nabla_i \tilde{\mu}_\N
\end{multline}
It is also worth noting that 
\begin{equation}
n_\n \en = n_\N \ec
\label{entrel}\end{equation}

The charged component equation includes both elastic and magnetic contributions.
The former can be written
\be
\Delta_\N f_i^\mathrm{el} = { 1 \over \rho_\N} \nabla^j  \sigma_{ij}
\ee
where
\be
\sigma_{ij} = \mu ( \nabla_i \xi^\N_j + \nabla_j \xi^\N_i) - { 2 \over 3} \mu (\nabla^l \xi_l^\N) \delta_{ij}
\ee
(here one should not confuse the shear modulus $\mu$  with the chemical potentials $\mu_\x$)
Meanwhile, the magnetic term follows from the standard electromagnetic Lorentz
force. That is, in this case we have
\be
f_i^\mathrm{mag}= f_i^\mathrm{L} = \frac{1}{c \rho_\N} \epsilon_{ijk} J^j B^k
\label{lorentz0}
\ee
Eliminating the total current with the help of Amp\'ere's law (i.e.  $J^i = (c/4\pi) \epsilon^{ijk} \nabla_j B_k$),
this becomes
\be
f_i^\mathrm{L} = { B^j \over 4\pi  \rho_\N} (\nabla_j B_i - \nabla_i B_j )\
\label{lorentz1}
\ee
Working out the Lagrangian variation using \citep{magnetcfs}
\be
\Delta_\N \left( {B^j \over \rho_\N} \right) = 0
\label{ind1}
\ee
we arrive at
\be
\Delta_\N B^i = -B^i \nabla_j \xi^j_\N  
\ee
and
\be
\Delta_\N B_i = B_j \nabla_i \xi_\N^j - B_i (\nabla_j \xi_\N^j) + B^j \nabla_j \xi^\N_i
\ee
Finally, using
\be
\Delta_\N ( \nabla_j B_i) = \nabla_j ( \Delta_\N B_i) - B_l \nabla_j \nabla_i \xi_\N^l
\ee
we obtain from (\ref{lorentz1})
\be
\Delta_\N f_i^\mathrm{mag}  = {B^j \over 4 \pi \rho_\N} [ \nabla_j ( \Delta_\N B_i) - \nabla_i (\Delta_\N B_j)]
\ee
These are all the relations we need to solve the problem. As far as we are aware, this is the first time that the
perturbation problem for combined superfluidity, elasticity and magnetic fields has been formulated. The equations we have given can be
directly applied to studies of global mode oscillations of a superfluid neutron star with a crust.

For later convenience it is  useful
to note that we could equally well have worked with Eulerian variations. In fact, since $f_\mathrm{L}^i=0$ in the background configuration
we must have $\Delta_\N f^\mathrm{mag}_i =\delta f^\mathrm{mag}_i$. Moreover, one can show that in the case of an incompressible fluid and a uniform background field
(see below)
we have
\be
\Delta_\N f^\mathrm{mag}_i= \delta f_i^{\rm mag} = \vA^2 \left [ (\B^j \nabla_j )^2 \xi_i^\N
-   \B^j \B^l \nabla_i \nabla_l \xi_j^\N  \right ]
\label{fnorm}\ee
where ''hats'' denote unit vectors. We have also defined the  Alfv\'en wave velocity
\be
\vA^2 = \frac{B^2}{4\pi\rho_\N}
\ee
It is important to note that in the superfluid system the Alfv\'en velocity scales with the number density of charged nucleons, 
not the total baryon number density \citep{mendell}.


\subsection{Plane-wave analysis}

As a first step towards understanding the problem, let us consider the simple case of a uniform, non-rotating background.
For an incompressible model, we have
\be
\nabla_i \xi^i_\x = 0 \longrightarrow \Delta_\x \tilde{\mu}_\x = 0
\ee
Then the problem simplifies to (note that we will have $\delta \Phi = \nabla_i \Phi=0$ for a uniform background)
\be
(1- \varepsilon_\n) \partial_t^2 \xi_i^\n + \varepsilon_\n \partial_t^2 \xi_i^\N = 0
\label{eq1}\ee
\be
(1- \varepsilon_\N) \partial_t^2 \xi_i^\N + \varepsilon_\N \partial_t^2 \xi_i^\n = \Delta_\N f_i^\mathrm{el}
+ \Delta_\N f_i^\mathrm{mag}
\label{eq2}\ee
We also need
\be
\Delta_\N f_i^\mathrm{el}  = v^2_s \nabla^2 \xi_i^\N
\ee
where the shear velocity, $v_s$, is defined by
\be
v_s^2 =  { \mu \over \rho_\N}
\label{shear}\ee
Note that the shear velocity scales with the number density of nucleons locked in the crust lattice, not the total nucleon number density as would be the case if there 
were no superfluid component. This distinction has not been made in previous work where the crust is modelled as a single component, see for example \citet{duncan,piro}.

We now consider (short wavelength $\ll$ the radius of the star) wave-propagation in this system. Making the standard plane-wave Ansatz
(see \citet{trev} for a recent analysis of the analogous non-magnetic two-fluid problem)
\be
\xi_i^\x = A_i^\x e^{i(\omega t + k_j x^j)}
\ee
where the index x is either n or c,
we have
\be
k^i A^\x_i = 0
\ee
i.e., the waves are transverse,
and
\be
\Delta_\N f_i^{\rm mag} = - \vA^2 k^2 \left [ (\B^j \hat{k}_j )^2 \xi_i^\N
- \hat{k}_i  ( \B_j \xi^j_\N ) ( \B^l \hat{k}_l )  \right ]
\label{fnorm}
\ee
From Eq.~(\ref{eq1}) we  immediately get the relation
\be
A_i^\n = - { \varepsilon_\n \over 1 - \varepsilon_\n} A_i^\N
\label{amprel}\ee
Using this in Eq.~(\ref{eq2}) we arrive at
\begin{multline}
\left[ { \omega^2 \over \varepsilon_\star}  - v^2_s k^2  -
\vA^2 k^2  (\hat{B}_j \hat{k}^j)^2 \right] A_i^\N \\
= - \vA^2 k^2 (\hat{B}_j \hat{k}^j)
(\B^l A_l^\N) \hat{k}_i
\label{chargeq}\end{multline}
where we have introduced
\be
\varepsilon_\star = { 1 - \varepsilon_\n \over 1 - \varepsilon_\n -  \varepsilon_\N }
\ee
Defining
\be
\omega_0^2 = v^2_s  k^2
\ee
the frequency of ``pure'' elastic waves, and the Alfv\'en-wave frequency
\be
 \omega_\mathrm{A}^2 = \vA^2 k^2
\ee
we have an equation for $A_i^\N$,
\be
\left[ { \omega^2 \over \varepsilon_\star} - \omega_0^2 -
 (\hat{B}_j \hat{k}^j)^2 \omega_\mathrm{A}^2  \right] A_i^\N = - \omega_\mathrm{A}^2  (\hat{B}_j \hat{k}^j)
(\hat{B}^l A_l^\N) \hat{k}_i
\label{ampli}
\ee
In order to arrive at the
final dispersion relation we first note that contracting the above equation
with $\hat{k}^i$ leads to the constraint
\be
\omega_\mathrm{A}^2  (\hat{B}_j \hat{k}^j) (\hat{B}^l A_l^\N) = 0
\ee
Thus, we can either choose to look for solutions
where $k^i$ is orthogonal to the local magnetic field or we see that
the polarisation $A_i^\N$, and hence $A_i^\n$, must be orthogonal to both
$k^i$ and $B^i$. Since the right hand side of (\ref{ampli}) vanishes in both cases we find
that all non-trivial solutions must be such that
\be
{ \omega^2 \over \varepsilon_\star} - \omega_0^2
- (\hat{B}_j \hat{k}^j)^2\omega_\mathrm{A}^2  = 0
\ee
That is, we have the general dispersion relation
\be
  \omega^2 = \omega_0^2 \varepsilon_\star \left[1 +
 (\hat{B}_j \hat{k}^j)^2 \frac{\omega_\mathrm{A}^2}{\omega_0^2} \right]
\label{frequ}\ee
Note that, in the
degenerate case when $B_i k^i=0$ we cannot uniquely determine the
polarisation; it can lie in any direction in the plane orthogonal to
$k^i$. Also, it is clear that such waves do not depend on the
magnetic field at all. Generically the polarisation is, however, well defined
(up to scale since we have a homogeneous system) to be orthogonal to
both $k^i$ and $B^i$.

We need to estimate the magnitude of the various terms. Let us first focus on the entrainment. Using (\ref{entrel}) 
we find that
\be
1 - \varepsilon_\n - \varepsilon_\N = 1 - { \varepsilon_\n \over x_\N}
\ee
where $x_\N = \rho_\N/\rho$. We can express this in terms of the effective
mass of the free neutrons, $m_\n^*$. Then we have (see \citet{prix} for a discussion of the analogous problem in a superfluid neutron star
core)
\be
\varepsilon_\n = 1 - { m_\n^* \over m_\n}
\ee
and it follows that
\be
\varepsilon_\star  = x_\N \left( 1 - x_\n {m_\n \over m_\n^*} \right)^{-1} = { x_\N \over \chi}
\ee
It is easy to show that $\chi^{-1/2}$ encodes the difference between the superfluid result and the standard result for a 
single component crust, i.e. with $\rho_\N \to \rho$ in (\ref{shear}).  
This may be the most meaningful comparison to make, since all previous studies of crust oscillations have 
assumed the single component model. 
 
\begin{figure}
\centering
\includegraphics[width=7cm,clip]{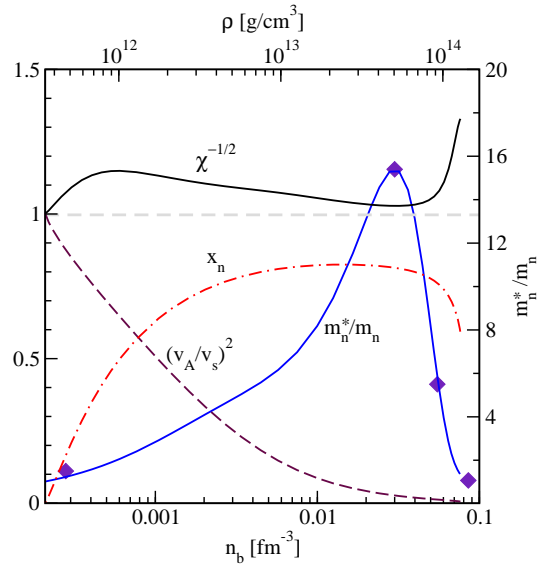}
\caption{This figure illustrates the density dependence of the different parameters that affect the wave propagation in the crust. 
We show, as functions of the total baryon number density $n_\mathrm{b}$, the superfluid neutron fraction $x_\n=n_\n/n_\mathrm{b}$ (dash-dot, left scale) for the equation of state 
discussed by \citet{douchin}, a fit to the effective neutron mass
$m_\n^*/m_\n$ (solid, right scale) based on the numerical results of \citet{nicolas} (data points indicated by diamonds), and the ratio between the Alfv\'en and the shear wave speeds 
$(v_\mathrm{A}/v_s)^2$ (dashed, left scale). The overall effect that the presence of the crust superfluid has on the local wave propagation, compared to the standard single component crust, 
is (as discussed in the main text) represented by $\chi^{-1/2}$ (solid, left scale). The horizontal dashed grey line indicates unity on the left scale.  }
\label{crustfig}
\end{figure}

What do we learn from these results? First of all, we see that in the limit $m_\n^* \to m_\n$, when the medium
effects that lead to the effective mass differing from the bare mass are not so great, we have
$\varepsilon_\star \to 1$ and $\chi \to x_\N$.  The waves in such a system, cf. (\ref{frequ}), are the usual shear waves with a (as we will see later)
relatively small magnetic correction. Of course, the results could still differ significantly from the standard single component model. The largest effect that one would expect 
would be, for $x_\N\approx 0.8$ cf. Figure~\ref{crustfig}, a frequency increase by about a factor of 2. However, the effective mass is expected 
to be larger than the bare mass so let us consider the  opposite limit, which may well apply in parts of the neutron star crust
[see, for example, \citet{nicolas}]. Then we have $m_\n^* \gg m_\n$. Using also $x_\N\le1$, we see that
$\varepsilon_\star \approx x_\N$ or $\chi \to 1$.
In this limit, it is very difficult for the free neutrons to move relative to the crust.
The upshot of this is that the waves in the system tend to the frequency predicted for 
a pure elastic crust \underline{without} a superfluid component.

These two extremes show that the presence of the superfluid in the neutron star crust can have a
significant effect on the frequencies of waves in the system. 
According to the data in Figure~\ref{crustfig}, the combined effect is at the 10\% level (compared to the single component crust result).
The results clearly show that
the superfluid component must be considered if we want to develop high precision magnetar crust seismology.
Of course, in reality we are mainly interested  in the global oscillations. Then the local effects that we have worked out will be
(in some sense) averaged throughout the crust. One may expect this to decrease the role of the superfluid since the effective neutron mass
may only be large in parts of the crust. Of course, the real answer requires a detailed mode calculation.
This problem remains to be solved. In order to provide reliable results, such an effort should draw
on more complete studies of the entrainment for the crust superfluid. One should also worry about the relevance
of vortex pinning and the mutual friction.

Finally, let us discuss the relative importance of the magnetic field. Scaling to "typical" values, we have
\be
\mu \approx 10^{13} \left( \frac{v_s}{10^8\, \mathrm{cm/s}} \right)^2 \left( \frac{\rho}{10^{14} \mathrm{g/cm}^3}
\right)\, \mathrm{dyne/cm}^2
\ee
 Then, it follows from (\ref{frequ}) that we need to consider
\be
\left( { \omega_A \over \omega_0} \right)^2 \approx 0.08
\left( {v_s \over 10^8 \mathrm{cm/s}}  \right)^{-2} \left( { \rho \over 10^{14} \mathrm{g/cm}^3} \right)^{-1}
\left( {B \over 10^{15}\ \mathrm{G}} \right)^2
\label{comp1}\ee
This shows that we can safely ignore the magnetic effects in the high-density region of the crust, cf. Figure~\ref{crustfig}. In order for the
magnetic term to dominate at the base of the crust we need $B \sim 10^{16}$~G,  stronger than the field strength
inferred for magnetars. Of course, one has to be a little bit careful here. First of all, it is entirely possible that the interior field
is stronger than the exterior dipole field which leads to the observed braking of the magnetar spin.
Secondly, (\ref{comp1}) indicates that the magnetic terms will dominate as we approach the surface of the star.
However, our analysis breaks down completely in the surface region. Basically, the MHD approximation is only valid as long as
the Alfv\'en wave speed is significantly below the speed of light. In order for this to be the case, we require
\be
\rho \gg 10^{8} \left( {B \over 10^{15}\ \mathrm{G}} \right)^2 \mathrm{g/cm}^3
\label{rhc}
\ee
When this is no longer true, as it will not be in a transition region near the surface of any neutron star, one must consider
the complete Maxwell equations. It is interesting we compare (\ref{rhc}) to the estimated density for the top of the crust, cf. for example
Eq.~(1) from \citet{piro}, 
\be
\rho_\mathrm{top} \approx 2.3 \times10^9 \left( {T \over 3\times10^8\ \mathrm{K} }\right)^3 \left({ 26 \over Z} \right)^6 \left({ A \over 56} \right) \mathrm{g/cm}^3
\ee
where $A$ is the number of nucleons  and $Z$ is the charge per ion, respectively. This density is scaled to iron, which means that the fiducial values should be relevant near the top of 
the crust. This suggests that MHD remains a valid approximation throughout a magnetar crust. The transition to electromagnetism takes place in the neutron 
stars envelope.  The nature of this transition is an important problem that deserves more attention.

\section{Multifluid dynamics of the core}

\subsection{Lagrangian perturbation equations}

The magnetohydrodynamics in the core is, in the simplest case, formulated in terms of three distinct fluids
associated with the neutrons, protons and electrons. The former two particle species are expected to be superfluid and
superconducting, respectively. Due to the smallness of the electron mass, the electron fluid degree of freedom can be suppressed and
the MHD equations effectively lead to a two-fluid model \citep{mendell}. The protons in the outer core are expected to form a type II
superconductor \citep{baym} which means that the magnetic field is carried by a large number of fluxtubes, each with a flux quantum $\phi_0 =  hc/2e$.
This should be the case provided the magnetic field is below a critical value $H_{\N 2} \approx 3 \times 10^{16}\, {\rm G} $
\citep{baym}. This critical value represents the field strength at which the magnetic fluxtubes overlap and can no longer be
treated individually. Above this threshold the magnetic field behaves "classically".
It should be noted that, even though the critical field is large, this is not an unrealistic possibility for magnetars
given that the magnetic field in the interior could be considerably higher than the exterior  dipole field.
Hence, one ought to consider both superconducting and normal protons. The latter case is, however, trivial. The desired result
follows immediately from the previous section, e.g. (\ref{frequ}), if we take the limit $\mu \to 0$.

For a non-rotating star, we again neglect the vortex-mediated mutual friction and the neutron vortex tension.
Omitting also a small entrainment induced magnetic term that originates from the London field by means of which the proton superconductor
rotates [see \citet{supercon} for discussion], the superfluid neutron dynamics is still governed by (\ref{eulern1}).

The combined proton-electron dynamics is a little bit more complicated. As discussed by \citet{supercon} the relevant equation
of motion
takes the form
\begin{multline}
(\partial_t + v^j_\N \nabla_j )\, (v_i^\N + \varepsilon_\N w_i^{\n \N} ) + \nabla_i( \tilde{\mu} +\Phi )
+ \varepsilon_\N w^{\n \N}_j \nabla_i v^j_\N \\
=  { 1 \over \rho_\N} \left( f^{\rm L}_i +  t_i^\N \right)
\label{eulerp0}
\end{multline}
where $v_\N^i$ and $v_\n^i$ are the velocites and $w_{\n\N}^i=v_\n^i-v_\N^i$, and we have (again) neglected the mutual friction.
There are two force terms on the right-hand side of this equation. The first, $f^{i}_{\rm L} $, is
the usual electromagnetic Lorentz force given by (\ref{lorentz0}). The second,  $t_\N^i $, represents the
smooth-averaged tension of the magnetic fluxtubes.
Remarkably, the Lorentz force does not play a role in the final superconducting MHD equations. As discussed by \citet{supercon} [see also \citep{mendell}]
it is
exactly cancelled by a term originating from the fluxtube tension. In the case of a non-rotating star
this leads to the magnetic force taking the form
\be
 f^{\rm L}_i +  t_i^\N = \frac{m_\N c}{4\pi e} \mathcal{W}^j_\N \left [ \nabla_j ( H_{\N 1} \hat{{\cal W}}^\N_i )
- \nabla_i ( H_{\N 1} \hat{{\cal W}}^\N_j )  \right ]
\label{tension}
\ee
where the lower critical magnetic field $H_{\N 1} = H_{\N 1}(\rho_\N) \approx 10^{15}$~G.

We have defined the vector ${\cal W}^i_\N $,
representing the (averaged) canonical proton vorticity \citep{prixmag}. This means that we have\footnote{Strictly speaking, this expression is
only valid for the non-rotating background. Perturbations of  ${\cal W}^i_\N $ will also contain the perturbed fluid velocities.
However, as long as we are focussing on the leading order contributions these can be neglected. That this is a legitimate approximation
is easy to see since
the characteristic frequency
\begin{displaymath}
{e \over m_\N c} B \approx 10^{19} \left ( \frac{B}{10^{15}\,{\rm G}} \right )\, {\rm s}^{-1}
\end{displaymath}
is much higher than any other relevant frequency in the problem.}
\be
{\cal W}^i_\N  \approx { e \over m_\N c}  B^i
\label{vort}
\ee
For a uniform incompressible model, this means that
\be
\delta f_i^\mathrm{mag} = { 1 \over \rho_\N} \delta ( f^{\rm L}_i +  t_i^\N)
\approx \frac{H_{\N 1}}{4\pi\rho_\N} B^j \left [\nabla_j \delta \hat{B}_i  - \nabla_i \delta \hat{B}_j \right ]
\label{fsu2}
\ee
In order to close the system of the MHD equations we need to provide a relation between
the magnetic field and the fluid velocity. This relation follows from the magnetic
induction equation. Neglecting the coupling forces between the electrons and
the neutron and proton fluids (see \citet{supercon} for discussion) the induction equation takes
the standard form,
\be
\partial_t B^i \approx \epsilon^{ijk} \epsilon_{klm} \nabla_j ( v^l_\N B^m)
\label{ind0}
\ee
Its perturbed form is given by (\ref{ind1}) from which we obtain
the Eulerian perturbation of the magnetic field,
\be
\delta B^i =  B^j \nabla_j \xi^i_\N - \nabla_j (\xi^j_\N B^i)
\label{dB}
\ee
Using this result, we find (for a uniform background and incompressible perturbations)
\be
\delta f^i_{\rm mag} \approx \cA^2  \hat{B}^j \hat{B}^l ( g^{ik} -\hat{B}^i \hat{B}^k   )
\nabla_j \nabla_l \xi^\N_k
\ee
where
\be
\cA^2 = \frac{H_{\N 1} B}{4\pi\rho_\N}
\ee
We now have all the relations we need to discuss short wavelength waves in the superfluid/superconducting system.


\subsection{Plane-wave analysis}

Most of the analysis works out exactly as in the crust problem (obviously in the
$\mu \to 0$ limit). The only difference is the form of the perturbed magnetic force.
With the plane-wave assumption, we see that
\be
\Dp f^{\rm mag}_i =  - \cA^2 k^2 (\hat{B}^j \hat{k}_j)^2 \left[  \xi_i^\N  -(\hat{B}^l \xi_l^\N)  \hat{B}_i \right]
\label{fsu3}
\ee
Comparing the magnetic forces in the normal and superconducting cases (equations (\ref{fnorm}) and (\ref{fsu3}),
respectively) we note some important differences. The characteristic speeds are
obviously different, $ \cA^2 = (H_{\N 1}/B) \vA^2 $. This is a well-known effect \citep{easson77,mendell}.
The two velocities would differ by
a factor $\sim 10^3 $ for a canonical pulsar with $ B = 10^{12}\, {\rm G} $. However, the effect will not be as dramatic
for magnetars for which $ B \approx H_{\N 1} \approx 10^{15}\, {\rm G} $.
Another difference arises from the second term in (\ref{fsu3}). This term is present
provided $\hat{B}^l \xi_l^\N \neq 0 $, or in other words when the wave-vector $k^i $ does not point along the direction of the magnetic
field. As a result, the magnetic force for superconducting protons has a non-zero component along
the magnetic field. Meanwhile, we have $ B_i \Dp f^i_{\rm mag} =0 $ in the normal proton case, cf. (\ref{fnorm}).
In the plane-wave problem the terms that give rise to this difference vanish identically. However, this
 rather subtle
difference in the magnetic forces will be relevant in many less idealised situations, e.g. for global mode oscillations.
Hence,  one must be careful before using intuition gained from standard MHD
problems in the case of a superconducting core. There is certainly more\footnote{For an interesting recent discussion on 
how superconductivity may affect the stability
properties of the star, see \citet{wasser}.} to the problem than a
simple "replacement" $B^2 \to B H_{\N 1}$ . 

Combining the perturbation equations as in the previous section, we readily arrive at
\be
\left [ { \omega^2 \over \varepsilon_\star}  -  \cA^2 k^2 (\hat{k}_j \B^j)^2 \right ]  A^i_\N =
 - \cA^2  k^2 (\hat{k}_j \B^j)^2 (\B_j A^j_\N ) \B^i
\label{equ1}
\ee
Working things out as in the crust case, we project this equation onto $\hat{k}_i$. Since 
$k_i A_\N^i=0$ it then follows that we have the constraint
\be
\cA^2 (\hat{k}_j \B^j)^3 (\B_j A^j_\N ) = 0
\ee
This has the same implications as in the crust problem. It immediately follows that the right-hand side of (\ref{equ1})
must vanish. Hence, we have the dispersion relation,
\be
\omega^2 = \varepsilon_\star\cA^2 k^2 (\hat{k}_j \B^j)^2 = \varepsilon_\star { H_{\N 1} B \over 4 \pi \rho_\N} k^2 (\hat{k}_j \B^j)^2
\label{freq2}
\ee

Our main interest here concerns the role of the superfluid neutron component. Its presence is reflected by the entrainment
factor in  (\ref{freq2}). To quantify its relevance, we express the entrainment in terms of
the effective proton mass, i.e. we use $\ec = 1 - m^*_\p/m_\p$. Then it follows, since the
proton fraction in the core is small, that
\be
\omega^2 \approx { m_\p \over m_\p^*} \cA^2 k^2 
\ee
Since it is expected that $0.3 <  m^*_\p/m_\p < 0.7 $ (see \citet{prix} for discussion) we see that the presence of the
superfluid neutrons will lead to a $\sim 20 - 80 \% $ {\it increase} in the frequency of the core waves.
This effect is large enough that cannot be neglected. It may, in fact, also be observable. If one accepts the
argument that the magnetic field couples motion in the crust to the core, and that the core fluid is
therefore partaking on the oscillation, then the entrainment will affect the observed frequencies.

We cannot at this point say much about the global oscillations of a magnetic
neutron star core; it is a problem that remains to be solved in detail. It is complicated
by the likely presence of an "Alfv\'en continuum" \citep{levin}. At this point it is not clear
to what extent the continuum
prevails in more detailed neutron star models. However, it is easy to see how the presence of the superfluid
will manifest itself in the continuum toy-model considered by \citet{levin}. The frequency range of the
continuum will simply scale according to (\ref{freq2}).

\section{Concluding remarks}

In this paper we have investigated the role of neutron star superfluidity for magnetar oscillations.
The results impact on attempts to use data from observed quasiperiodic oscillations in the tails of magnetar flares
to place contraints on neutron star parameters, see for example \citet{lars}. 
Using a plane-wave analysis we  estimated the effects of the
neutron superfluid in the elastic crust region. This, the first ever, analysis of the combined magnetic-elastic-superfluid
crust problem demonstrated that  the superfluid
imprint is likely to  be more significant than the effects of the crustal magnetic field. This is, of course, assuming that the 
SGR flare mechanism does not deposit sufficient heat in the crust to raise the system above the superfluid transition temperature. 
Available estimates, e.g. \citet{lyb}, suggest that this is unlikely. 
We also considered the region immediately beneath the crust, where superfluid neutrons are thought
to coexist with a type~II proton superconductor. Since the magnetic field in the latter is
carried by an array of fluxtubes, see \citet{supercon} for a recent discussion, the dynamics of this region differs from standard magnetohydrodynamics.
We showed that the presence of the neutron superfluid (again) affects the oscillations of the system. This accords well with previous results of, in particular,
\citet{mendell}.

Our estimates show that the superfluid components cannot be ignored
in efforts to carry out magnetar seismology. This increases the level of complexity of the
modelling problem, but also points to the exciting possibility of using 
observations to probe the superfluid nature of
supranuclear matter. Future work needs to extend our analysis to consider the global oscillations of 
magnetic-superfluid-elastic neutron stars. This is a very interesting problem because, in addition to enabling a more detailed
seimsology analysis, it may also provide insight into rotational glitches in magnetars \citep{dib}. It is generally believed that
superfluidity plays a key role in radio pulsar glitches. Recent developments in modelling these events is showing some promise \citep{glitch},
and it would obviously be highly relevant to extend this analysis to strongly magnetised systems.


\section*{Acknowledgements}
We would like to thank Anna Watts for helpful discussions.
This work was supported by STFC in the UK through grant number PP/E001025/1.
KG is supported by the German Science Foundation (DFG) via SFB/TR7.

\end{document}